# Optical Kerr nonlinearity and multi-photon absorption of DSTMS measured by Z-scan method

**JIANG LI,**[1,2,3,4] **RAKESH RANA,**[1] **LIGUO ZHU,**[2,3] **CANGLI LIU,**[3] **HARALD SCHNEIDER,**[1] **AND ALEXEJ PASHKIN**[1,*]

[1]*Institute of Ion Beam Physics and Materials Research, Helmholtz-Zentrum Dresden-Rossendorf, Dresden, 01328, Germany*
[2]*Institute of Fluid Physics, China Academy of Engineering Physics, Mianyan, Sichuan 621000, China*
[3]*Microsystem & Terahertz Research Center, China Academy of Engineering Physics, Chengdu, Sichuan 610200, China*
[4]*tjujiangli@tju.edu.cn*
*\*o.pashkin@hzdr.de*

**Abstract:** We investigate the optical Kerr nonlinearity and multi-photon absorption (MPA) properties of 4-N, N-dimethylamino-4'-N'-methyl-stilbazolium 2, 4, 6- trimethylbenzene-sulfonate (DSTMS) excited by femtosecond pulses at a wavelength of 1.43 μm, which is optimal for terahertz generation via difference frequency mixing. The MPA and the optical Kerr coefficients of DSTMS at 1.43 μm are strongly anisotropic indicating a dominating contribution from cascaded 2nd-order nonlinearity.

## 1. Introduction

Single-cycle THz pulses with high electric field amplitudes have recently attracted large attention since they provide novel opportunities in photonics [1], condensed matter physics [2], charged-particle accelerators [3], surface chemistry [4] and biomedical science[5]. The advent of transient high-field pulses reaching multiple MV/cm that oscillate at frequencies in the range from 1 THz to 10 THz has provided an emerging tool to excite extremely non-equilibrium dynamic phenomena by driving phonons or other low-energy collective modes [6]. The most prominent laser-based approaches providing such intense THz pulses transients at MV/cm field strength are based on nonlinearities in plasma [7], optical rectification (OR) in lithium niobate (LN) [8, 9], and OR in nonlinear organic crystals (NOCs) [10–15]. Compared with the other two methods, OR in organic crystals offers several advantages: (i) NOCs, including DAST, DSTMS, OH1, HMQ-TMS, possess a large 2nd-order nonlinear optical susceptibility at room temperature (a comparison for typical nonlinear crystals for THz generation are shown in Table 1), resulting in high THz conversion efficiency; (ii) phase matching can be achieved in a collinear geometry and it does not require sophisticated pump-pulse shaping; (iii) the THz radiation is naturally collimated and aberration-free, which makes it possible to focus the beam by a single optics to a diffraction-limited spot and to reach high field strength. Among NOCs, 4-N, N-dimethylamino-4'-N'-methyl-stilbazolium 2, 4, 6- trimethylbenzene-sulfonate (DSTMS) provides very broadband phase matching conditions with a minor phonon absorption and can generate octave-spanning spectra up to 26 THz [17], resulting in short pulses with ultra-high peak fields [11–14]. Since the quest for higher field strength is incessant, upscaling of the THz pulse energy calls for increasing pump fluence. However, the THz conversion efficiency shows strong saturation effects upon intense femtosecond (*f*s) pulse pumping in DSTMS [11]. To figure out the main limitation factors, it is critical to know high-order nonlinear properties of DSTMS going beyond its 2nd-order nonlinear optical susceptibility. The most important of them are the Kerr nonlinearity and multi-photon absorption (MPA).

In this article, the Z-scan method is implemented to measure the polarization-dependent nonlinear refractive index (Kerr nonlinearity) and the MPA coefficient of DSTMS at the wavelength of 1.43 μm, which is close to optimal phase matching condition for THz generation. The experimental results show that both MPA and the optical Kerr nonlinearity possess high anisotropy. The optical Kerr nonlinearity of DSTMS at 1.43 μm is dominated by the cascaded second-order nonlinear effect, while its intrinsic third-order nonlinearity is very weak.

Table 1. Selection of organic and inorganic nonlinear crystals for THz generation and their most relevant parameters: nonlinear coefficient $d_{eff}$, spectral range of the most efficient THz emission, damage threshold, the figure of merit $FM_{THz} = d_{eff}^2/n_0^2 n_{THz}$ for THz generation [16]. The THz spectral ranges are adopted from experimental results in Ref. [8, 14,17-19], and the damage thresholds are quoted for the optimal phase-matching wavelength in Ref. [20-24]. The nonlinear organic crystals survive higher peak powers due to the weaker multi-photon absorption at the optimal operation wavelength

| Nonlinear crystals | $d_{eff}$ (pm/V) | Spectral range (THz) | Damage threshold (GW/cm$^2$) | $FM_{THz}$ (pm/V)$^2$ | Wavelength of optimal phase matching (μm) |
|---|---|---|---|---|---|
| DAST | 240 | 0.1-10 | 300 | 5600 | 1.5 |
| DSTMS | 230 | 0.5-26 | 150 | 5800 | 1.5 |
| ZnTe | 66 | 0.1-3 | 100 | 170 | 0.8 |
| GaP | 24 | 0.3-7 | 66.2 | 17 | 1.0 |
| LiNbO$_3$ | 160 | 0.1-2.5 | 12200 | 1100 | 1.0 |

## 2. MATERIAL CHARACTERIZATION

For organic molecular crystal, the nonlinear optical response is decided by molecular symmetry. In centrosymmetric molecular crystals, the two-photon absorption (2PA) may be forbidden by the symmetry even though the photon energy is sufficient for the 2PA due to selections rule based on the parity of the electronic wavefunctions [25]. Since DSTMS is non-centrosymmetric, it has similar nonlinear optical responses, compared to semiconductors. The optical Kerr nonlinearity and the MPA process is determined by the dispersion of the linear and nonlinear refractive index [26], which is strongly dependent on semiconductor bandgap. To estimate the polarization dependence of the bandgap in DSTMS, we have measured absorption spectra using a FTIR spectrometer (Vertex 80v, Bruker Optics Corp.). The studied sample was a commercial z-cut DSTMS crystal with a thickness of 0.4 mm (Rainbow Corp.). The bandgap for the light polarization along the two nearly orthogonal crystallographic directions ([100] and [010]) is obtained from the Tauc plot of measured visible absorption spectra, as depicted in Fig. 1(a). The bandgap of DSTMS shows a slight anisotropy, resulting in the blue shift of the absorption, when rotating the light polarization from the [100] to the [010] axis. The total bandgap variation shown in Fig. 1(b) is less than 2%.

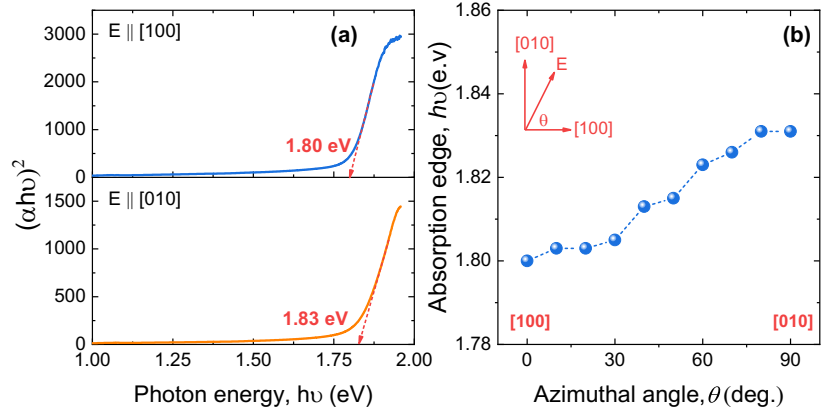

Fig. 1. (a) Tauc plot of absorption spectra for the light polarization along the [100] and [010] axes. (b) azimutal angle-dependent absorption edge of DSTMS, here θ is the angle between [100] and the light polarization.

## 3. Z-SCAN MEASUREMENT

The scheme of the Z-scan experiment is shown in Fig. 2. An optical parametric amplifier (OPA) seeded by a Ti: sapphire amplifier serves as a source of *fs* pulses with a center wavelength of 1.43 μm with a duration of 75 *fs* FWHM at a repetition rate of 1 kHz. The linearly polarized OPA radiation passes through a λ/2 waveplate to control the polarization plane and then it is focused by the lens $L_1$ with a focal length of 150 mm. The measured focused spot size was 22 μm ($1/e^2$ intensity radius), and the pulse energy was 20 nJ, corresponding to an incoming on-axis peak intensity of $I_{in}$ = 32.9 GW/cm$^2$.

The DSTMS crystal is mounted on a motorized linear translation stage (Newport Corp.) with a scanning resolution of 5 μm. The beam transmitted through the sample and the aperture is collimated by the lens $L_2$ and, after a long-pass (LP) filter (Si wafer, to block fluorescence due to MPA process) it is focused on a InGaAs photodiode (PD). A lock-in amplifier is utilized for the acquisition of the signal from the PD. The Z-scan records the transmittance with a fully open aperture (OA), $T_{OA}$, and the transmittance with a partially closed aperture (CA), $T_{CA}$ as a function of the sample position along the beam axis. For all scans, the aperture size for CA Z-scans is set in a low-fluence limit to the transmittance level of S=0.25. To eliminate scattering artefacts due to the surface roughness and imperfections in the DSTMS crystal, both $T_{OA}$ and $T_{CA}$ traces were subtracted by OA and CA Z-scan curves measured at low fluence. The OA Z-scan trace characterizes multi-photon absorption, whereas the nonlinear refractive index ($n_2$) can be estimated from the normalized transmittance trace, $T_{CA}/T_{OA}$.

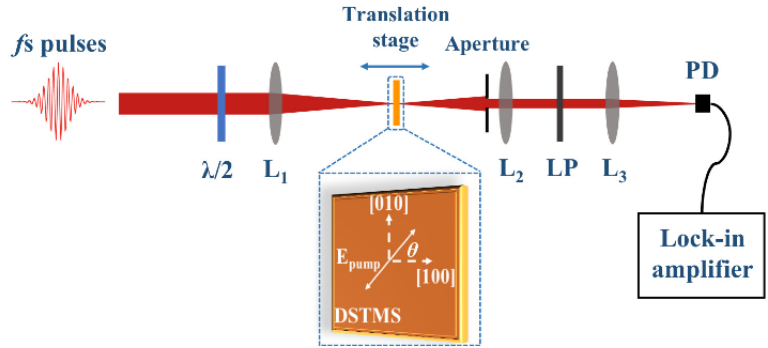

Fig. 2. Scheme of Z-scan set-up. $L_1$, $L_2$, $L_3$: lens; PD: photodiode; LP: long-pass filter for 1.43 μm. θ is the angle between polarization of *fs* pulses and [100] of DSTMS, as shown in insert

## 4. RESULTS AND DISCUSSION

The measured normalized transmittances of OA and CA with polarization of *f*s pulses along the [100] (blue solid dots) and [010] (red solid dots) directions are shown in Fig. 3. The optical Kerr nonlinearity and MPA demonstrate a clear anisotropy. If only one type of MPA dominates for a given wavelength, the optical intensity, $I(z, R, t)$ can be described by,

$$\frac{dI(z,R,t)}{dz} = -\alpha_N I^N(z, R, t) \tag{1}$$

where $z$ is the propagation distance, $R$ is the transverse coordinate, $t$ is time, and $\alpha_N$ is the N-photon absorption coefficient. Relative change of transmission $(T - T_0)/T_0 \ll 1$, Eq. 1 results in [27],

$$\frac{T(z)}{T_0} = \frac{1}{1+N^{-3/2}\alpha_N(I_0/(1+z^2/z_0^2))^{N-1}l} \tag{2}$$

where $l$ is the thickness of a NOC, $I_0$ is the peak on-axis intensity at focus, $z_0$ is the Rayleigh length. By fitting with Eq. 2, MPA coefficients can be extracted from the curves. Fig. 2(a) shows fits for two- and three-photon absorption (N =2 and 3, respectively). Clearly, the 3PA model gives the best fitting results for the wavelength of 1.43 μm. This is expected, since the doubled photon energy at this wavelength is 1.73 eV, which below the bandgap of DSTMS (see Fig. 1(a)).

In a typical Z-scan measurement, the normalized transmittance, $T_{CA}/T_{OA}$ at a given position of sample z is expressed as [28],

$$T(z) = \frac{T_{CA}}{T_{OA}} = \frac{\int_{-\infty}^{\infty} P_T(z,t)dt}{S\int_{-\infty}^{\infty} P_{in}(z,t)dt} \tag{3}$$

$$P_T(z,t) = c\varepsilon_0 n_0 \pi \int_0^{R_a} |E(z,R,t) \cdot exp(-\alpha_0 L/2 + i\Delta\Phi(z,R,t))|^2 RdR \tag{4}$$

$$\Delta\Phi(z, R, t) = \frac{\Delta\Phi_0(t)}{1+z^2/z_0^2} exp\left(-\frac{2R^2}{\omega^2(z)}\right) \tag{5}$$

where $P_T(z, t)$ is the transmitted power through the aperture, $c$ is velocity of light in vacuum, $\varepsilon_0$ is the permittivity of vacuum, $n_0$ is the linear index of refraction, $R_a$ is the radius of aperture; $S$ is the aperture transmittance in linear regime. $P_{in}(t) = \pi\omega_0^2 I_0(t)/2$ is instantaneous input power within the sample, $I_0(t)$ is the on-axis irradiance at focus, $\omega(z)$ is the beam radius, and $\omega_0 = \omega(0)$ is the beam radius at focus; $\Delta\Phi_0(t) = kn_2 I_0(t)L_{eff}$ is on-axis phase shift at focus, $k$ is wave vector of incident laser beam, $n_2$ is nonlinear refractive index of the sample, $L_{eff} = (1 - e^{-\alpha_0 l})/\alpha_0$ is the effective propagation length within sample, and $\alpha_0$ is the linear absorption coefficient with the measured value of about 2 cm$^{-1}$ in the wavelength range from 0.7 μm to 1.5 μm [29]. Since the estimated experimental phase shift, $\langle\Delta\Phi_0\rangle$ ($\langle\Delta\Phi_0\rangle \cong \Delta T_{p-v}/0.406(1-S)^{0.25}$, $\Delta T_{p-v}$ is the variant of peak-to-valley in $T_{CA}$ [28]) is smaller than $0.23\pi$, the normalized transmittance is determined in the limit of small nonlinear phase change [28],

$$T(z, \langle\Delta\Phi_0\rangle) = 1 - \frac{4\langle\Delta\Phi_0\rangle x}{(x^2+1)(x^2+9)} \tag{6}$$

where x = $z/z_0$, $\langle\Delta\Phi_0\rangle = \langle\Delta\Phi_0(0)\rangle/\sqrt{2}$ is the averaged instantaneous phase $\Delta\Phi_0(t)$ over the laser pulse shape with a Gaussian pulse approximation. By fitting with Eq. 6, the sign and value of n$_2$ can be obtained from the measured results.

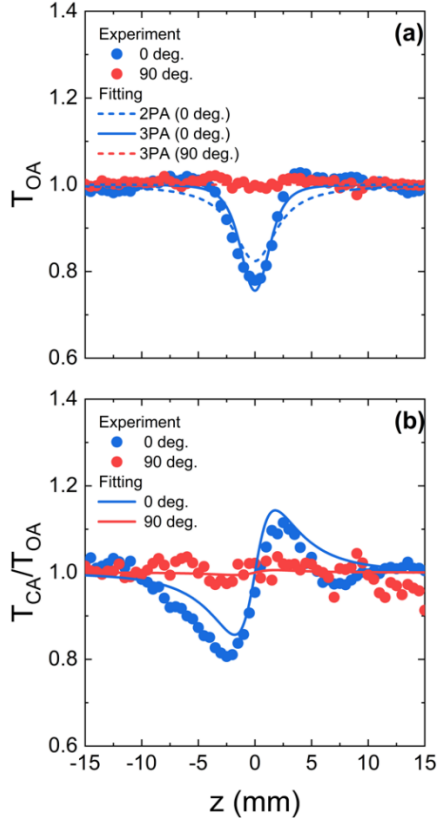

Fig. 3. Normalized transmittance with full open aperture (a) and relative transmittance with closed aperture (b) at wavelength of 1.43 μm.

To figure out the azimuthal angle dependence of the 3PA and the optical Kerr nonlinearity of DSTMS, normalized transmittances of OA and CA traces at different azimuthal angles were measured by rotating the polarization of incident *f*s pulses by the λ/2 waveplate. The *z*-cut DSTMS crystal has a large birefringence in the [001] plane with a difference in the group refractive index of $\Delta n_g \approx 0.5$ [29]. Thus, the orthogonal polarization components of the incident pulse become spatially separated already after propagating only 45 μm through the DSTMS crystal. This distance is much smaller than the total thickness of our sample (400 μm). Therefore, the contributions to the Z-scan signal of the polarization components along the [100] and [010] directions can be considered as nearly independent, i.e., the total signal can be calculated as a sum of both contributions. Moreover, although DSTMS has a large second-order nonlinear coefficient for THz generation, the THz conversion efficiency is below 1% at our pump fluence level. Hence, the loss caused by THz conversion is negligible in the CA measurements.

Fitting with Eq. 2 and Eq. 6 gives the 3PA coefficients ($\gamma$) and $n_2$ as a function of $\theta$. As shown in Fig. 4, both $\gamma$ and $n_2$ are highly anisotropic at 1.43 μm. When rotating the polarization of the electric field from [100] to [010], the value of $\gamma$ changes from $(6.02 \pm 0.51) \times 10^{-2}$ cm$^3$/GW$^2$ to near to zero (($0.01 \pm 0.14) \times 10^{-2}$ cm$^3$/GW$^2$). The errors are determined by 95 % confidence interval in the fitting. The $\cos^6\theta$ dependence is depicted in Fig. 4(a). To distinguish the highly anisotropic 3PA behavior further, the normalized transmittance, $T(0)/T_0$ ($T_0$ is the transmittance out of focus point), as a function of peak intensity at focus point was investigated, as shown in Fig. 5. For the polarization of optical pulses along to [100] axis, a significant drop in transmittance with increasing peak intensity can be observed, whereas the normalized transmittance keeps near to 1 with up to 100 GW/cm² when polarization along to [010] axis. The peak intensity dependent transmittance along to [010] axis was well reproduced by $T(0) = 1/\sqrt{1 + 2\gamma L_{eff} I_0^2}$ with the assumption of the 3PA dominating nonlinear absorption. The highly anisotropic 3PA coefficients is related to bound electron response in the organic DSTMS crystal. The crystalline structure of DSTMS is shown in inset of Fig. 4(a) (adopted from Ref. 30), the electric field parallel to the [100] direction (0-degree angle in the experiment) polarizes the molecules along their longest axes. In this direction some electronic orbitals can have a very large extension and, correspondingly, a large dipole moment naturally causing a stronger nonlinear response. The electric field parallel to the [010] direction (90 degrees) is applied either along the shorter molecular axes or almost normal to the molecular plane. Therefore, a weaker nonlinearity is expected in this direction.

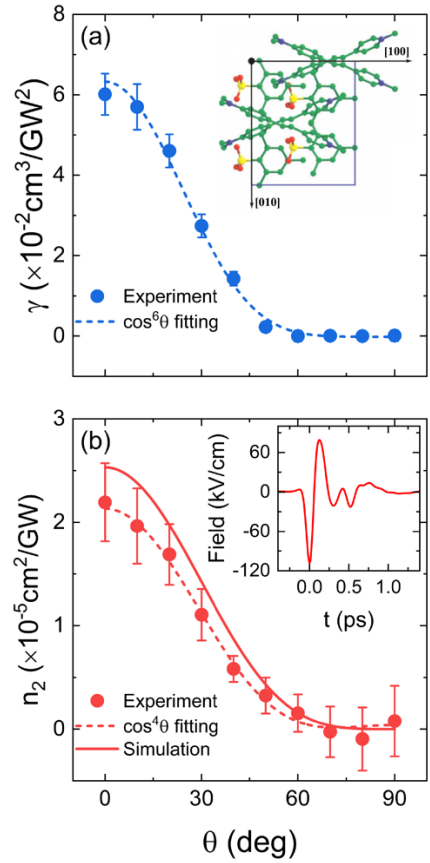

Fig. 4. Azimuthal angle dependent (a) 3PA coefficients and (b) nonlinear refractive index, $n_2$ at a wavelength of 1.43 μm. The inset in (a) shows the crystalline structure of DSTMS, and the insets in (b) shows the simulated THz-field in the time-domain with pump fluence of 2.6 mJ/cm² at 1.43 μm

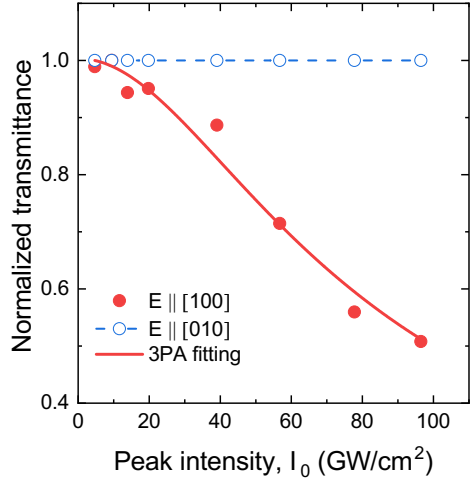

Fig. 5. Peak intensity dependent normalized transmittance at focus point at wavelength of 1.43 μm.

The nonlinear refractive index $n_2$ shows a similar behavior, decreasing from $(2.19 \pm 0.28) \times 10^{-5}$ cm/GW to $(0.08 \pm 0.34) \times 10^{-5}$ cm/GW, as shown in Fig. 4(b). In the case of a non-centrosymmetric crystal with a large $\chi^{(2)}$ nonlinearity such as DSTMS, there are several contributions to the total nonlinear refractive index [31],

$$n^{total} = n^{direct} + n^{SHG} + n^{OR} \qquad (7)$$

where $n^{direct} \propto Re\{\chi^{(3)}\}$ is the contribution from the intrinsic $\chi^{(3)}$ nonlinearity; $n^{SHG} \propto d_{eff}^2/\Delta k$ is the contribution from 2$^{nd}$-order cascaded processes duo to second-harmonic generation (SHG) [32], $d_{eff}$ is the effective nonlinear optical coefficient, $\Delta k = k^{2\omega} - k^{\omega}$ is the wave vector mismatch; $n^{OR} \propto r_{iik}^2$ is the contribution from 2$^{nd}$-order cascaded processes due to combination of optical rectification (OR) and the linear electro-optic (EO) effect [31], $r_{iik}$ is the electro-optic coefficient. When the incident $fs$ pulses propagate along the [001] axis, both type I and type II are far from phase matching for second-harmonic generation. Thus, $n^{SHG}$ should be negligible. On the other hand, DSTMS possesses a large nonlinear optical susceptibility $\chi^{(2)} = (430 \pm 40)$ pm/V and an electro-optic coefficient $r_{111} = (37 \pm 3)$ pm/V [29]. Moreover, the [100] light polarization fulfills the optimal THz generation conditions. Therefore, the contribution from $n^{OR}$ is maximum for the polarization along the [100] direction and it should have a $\cos^4(\theta)$ dependence on the azimuthal angle $\theta$. Therefore, the $n_2$ of DSTMS as a function of azimuthal angle is given by [33],

$$n^{total} \propto a_1 \cos^4(\theta) + a_2 \sin^4(\theta) + a_3 \frac{\sin^2(\theta)}{4}, \qquad (8)$$

$$a_1 = Re\{\chi^{(3)}_{1111}\} + n^{OR}, \qquad (9)$$

$$a_2 = Re\{\chi^{(3)}_{2222}\}, \qquad (10)$$

$$a_3 = Re\left\{\sum_{\text{off-diag}}^{(1,2)} \chi^{(3)}_{ijkl}\right\} \qquad (11)$$

Here, $a_1$ is equal to the sum of the intrinsic $\chi^{(3)}$ diagonal tensor component for the [100] axis and $n^{OR}$, $a_2$ and $a_3$ are given by the intrinsic $\chi^{(3)}$ diagonal tensor component for the [010] axis and the off-diagonal components, respectively. The line in Fig. 4(b) shows the best fit of the measured $n^2$ using Eq. 8. The contributions from the last two terms are close to zero ($a_2$ = (0.04 ± 0.06) × 10$^{-5}$ cm$^2$/GW, $a_3$ = (−0.43 ±0.19) × 10$^{-5}$ cm$^2$ /GW). Therefore, the combination of the intrinsic $\chi^{(3)}$ process along to [100] and $n^{OR}$, the value of which is $a_1$ = (2.13 ± 0.37) × 10$^{-5}$ cm$^2$ /GW, dominates the $n_2$.

To extract the contribution from the quasi-$\chi^{(3)}$ effect induced by the OR and EO effects, the approximations of plane waves and non-depleted pump are assumed. Considering THz-wave absorption and non-perfect phase matching, the generated electric field $E(\Omega, \lambda, L)$ at the angular THz frequency $\Omega$ pumped by an optical pulse with center wavelength $\lambda$ is given by [16],

$$E(\Omega, \lambda, L) = \frac{2d_{THz}\Omega^2 I(\Omega)}{(\Omega(n_{THz}+n_g)/c + i\alpha_{THz}/2)n\varepsilon_0 c^3} \times \frac{exp\{i(\Omega n_{THz}/c + i\alpha_{THz}/2)L\} - exp(i\Omega n_g L/c)}{\Omega(n_{THz}-n_g)/c + i\alpha_{THz}/2} \quad (12)$$

where $I(\Omega) = I_0 \tau exp(-\tau^2\Omega^2/2)$ for Gaussian pulse, $I_0$ is the peak intensity; $\varepsilon_0$, $c$ is the permittivity and light velocity in vacuum; $d_{THz}$ and $L$ are the nonlinear coefficient for THz generation and thickness of the DSTMS crystal; $n_{THz}$ and $\alpha_{THz}$ are the refractive index and absorption coefficient at THz angular frequency; $n$ and $n_g$ are refractive index and group refractive index at optical wavelength. By taking the Fourier transform of Eq. 12, the electric field of THz pulses in the time-domain can be obtained at the experimental pump fluence of 2.63 mJ/cm$^2$ at 1.43 μm, as shown in the inset of Fig. 4(b). The on-axis refractive index change induced by the THz field along [100] due to the EO effect is given by [34],

$$\Delta n^{OR} = -\frac{1}{2}n^3 r_{111}\langle E_{THz}^{peak}\rangle \quad (13)$$

$\langle E_{THz}^{peak}\rangle$ is the averaged peak electric field of THz pulses within DSTMS. $n^{OR}$ along the [100] direction is obtained by $n_2^{OR} = \Delta n^{OR}/I_0$, the corresponding value of which is 2.53 × 10$^{-5}$ cm$^2$/GW. As mentioned above, n$^{OR}(\theta)$ has a cos$^4(\theta)$ dependence. As Fig. 4(b) shows, the simulation results agree well with measured n$^{OR}(\theta)$, demonstrating that the nonlinear refractive index $n_2$ is mainly originating from the quasi-$\chi^{(3)}$ effect due to the OR and EO effect. The intrinsic $n_2$ at 1.43 μm should be negligible compared to this contribution. This assumption is confirmed by the Z-scan measurements carried out at the wavelength of 1.87 μm (the OPA idler signal). For this wavelength the THz generation is negligible due to bad phase matching condition and the measured signals for the optical Kerr nonlinearity and the MPA for all polarizations are very weak and below the sensitivity of our setup.

## 5. CONCLUSION

We have measured the optical Kerr nonlinearity and the 3PA coefficients of DSTMS at the wavelength of 1.43 μm using a single-beam Z-scan method. The experimental results indicate that the 3PA and the optical Kerr coefficients of DSTMS at 1.43 μm are anisotropic. The nonlinear refractive index $n_2$ is dominated by the cascaded 2$^{nd}$-order processes due to the optical rectification. The intrinsic optical Kerr nonlinearity is negligible compared to the cascaded process at the wavelength of 1.43 μm. This finding is further corroborated by the vanishingly small optical Kerr nonlinearity and MPA observed in comparative measurements at 1.87 μm excitation in the absence of efficient optical rectification.

**Funding.** Foundation of President of China Academy of Engineering Physics (Grant No. YZJJLX2018001), National Natural Science Foundation of China (Grant No.11704358, 12002326)

**Acknowledgments.** Jiang Li thanks China Scholarship Council (file no.201804890029) for financial supports.

**Disclosures.** The authors declare no conflicts of interest.

### References

1. S.-C. Chen, Z. Feng, J. Li, W. Tan, L.-H. Du, J. Cai, Y. Ma, K. He, H. Ding, Z.-H. Zhai, Z.-R. Li, C.-W. Qiu, X.-C. Zhang, and L.-G. Zhu, "Ghost spintronic THz-emitter-array microscope," Light Sci. Appl. 9, 99 (2020).
2. O. Schubert, M. Hohenleutner, F. Langer, B. Urbanek, C. Lange, U. Huttner, D. Golde, T. Meier, M. Kira, S. W. Koch, and R. Huber, "Sub-cycle control of terahertz high-harmonic generation by dynamical Bloch oscillations," Nat. Photonics 8, 119–123 (2014).

3. D. Zhang, A. Fallahi, M. Hemmer, X. Wu, M. Fakhari, Y. Hua, H. Cankaya, A. L. Calendron, L. E. Zapata, N. H. Matlis, and F. X. Kärtner, "Segmented terahertz electron accelerator and manipulator (STEAM)," Nat. Photonics 12, 336–342 (2018).
4. J. L. LaRue, T. Katayama, A. Lindenberg, A. S. Fisher, H. Öström, A. Nilsson, and H. Ogasawara, "THz-Pulse-Induced Selective Catalytic CO Oxidation on Ru," Phys. Rev. Lett. 115, 36103 (2015).
5. Y. Li, C. Chang, Z. Zhu, L. Sun, C. Fan, "Terahertz wave enhances permeability of the voltage-gated calcium channel," J. Am. Chem. Soc. 143(11), 4311-4318 (2021).
6. T. Kampfrath, K. Tanaka, and K. A. Nelson, "Resonant and nonresonant control over matter and light by intense terahertz transients," Nat. Photonics 7, 680–690 (2013).
7. M. Clerici, M. Peccianti, B. E. Schmidt, L. Caspani, M. Shalaby, M. Giguère, A. Lotti, A. Couairon, F. Légaré, T. Ozaki et al., "Wavelength scaling of terahertz generation by gas ionization," Phys. Rev. Lett. 110, 253901 (2013).
8. H. Hirori, A. Doi, F. Blanchard, and K. Tanaka, "Single-cycle terahertz pulses with amplitudes exceeding 1 MV/cm generated by optical rectification in $LiNbO_3$," Appl. Phys. Lett. 98, 091106 (2011).
9. K.-L. Yeh, M. Hoffmann, J. Hebling, and K. A. Nelson, "Generation of 10 μJ ultrashort terahertz pulses by optical rectification," Appl. Phys. Lett. 90, 171121 (2007).
10. C. P. Hauri, C. Ruchert, C. Vicario, and F. Ardana, "Strong-field single-cycle THz pulses generated in an organic crystal," Appl. Phys. Lett. 99, 161116 (2011).
11. C. Vicario, B. Monoszlai, and C. P. Hauri, "GV/m single-cycle terahertz fields from a laser-driven large-size partitioned organic crystal," Phys. Rev. Lett. 112, 213901 (2014).
12. C. Vicario, A. Ovchinnikov, S. Ashitkov, M. Agranat, V. Fortov, and C. Hauri, "Generation of 0.9-mJ THz pulses in DSTMS pumped by a $Cr:Mg_2SiO_4$ laser," Opt. Lett. 39, 6632–6635 (2014).
13. M. Shalaby and C. P. Hauri, "Demonstration of a low-frequency three-dimensional terahertz bullet with extreme brightness," Nat. Commun. 6, 1–8 (2015).
14. C. Vicario, M. Jazbinsek, A. Ovchinnikov, O. Chefonov, S. Ashitkov, M. Agranat, and C. Hauri, "High efficiency thz generation in DSTMS, DAST and OH1 pumped by Cr:forsterite laser," Opt. Express 23, 4573–4580 (2015).
15. T. O. Buchmann, E. J. Kelleher, K. J. Kaltenecker, B. Zhou, S.-H. Lee, O.-P. Kwon, M. Jazbinsek, F. Rotermund, and P. U. Jepsen, "MHz-repetition-rate, sub-mW, multi-octave THz wave generation in HMQ-TMS," Opt. Express 28, 9631–9641 (2020).
16. M. Jazbinsek, U. Puc, A. Abina, and A. Zidansek, "Organic crystals for THz photonics," Appl. Sci. 9, 882 (2019).
17. C. Somma, G. Folpini, J. Gupta, K. Reimann, M. Woerner, and T. El-saesser, "Ultra-broadband terahertz pulses generated in the organic crystal DSTMS," Opt. Lett. 40, 3404–3407 (2015).
18. F. Blanchard, L. Razzari, H.-C. Bandulet, G. Sharma, R. Morandotti, J.-C. Kieffer, T. Ozaki, M. Reid, H. F. Tiedje, H. K. Haugen, and F. A. Hegmann, "Generation of 1.5 μJ single-cycle terahertz pulses by optical rectification from a large aperture ZnTe crystal," Opt. Express 15, 13212-13220 (2007).
19. K. Aoki, J. Savolainen, and M. Havenith, "Broadband terahertz pulse generation by optical rectification in GaP crystals," Appl. Phys. Lett. 110, 201103 (2017).
20. B. Monoszlai, C. Vicario, M. Jazbinsek, and C. P. Hauri, "High-energy terahertz pulses from organic crystals: DAST and DSTMS pumped at Ti:sapphire wavelength," Opt. Lett. 38, 5106-5109 (2013).
21. B. Liu, H. Bromberger, A. Cartella, T. Gebert, M. Först, and A. Cavalleri, "Generation of narrowband, high-intensity, carrier-envelope phase-stable pulses tunable between 4 and 18 THz," Opt. Lett. 42, 129-131 (2017).
22. T. Yuan, J.Z. Xu, and X.-C. Zhang, "Development of terahertz wave microscopes," Infrared Phys. Techn. 45, 417–425 (2004).
23. Y. Li, F. Liu, Y. F. Li, L. Chai, Q. R. Xing, M. L. Hu, and C. Y. Wang, "Experimental study on GaP surface damage threshold induced by a high repetition rate femtosecond laser," Appl. Opt. 50, 1958-1962 (2011).
24. Z. L. Su, Q. L. Meng, and B. Zhang, "Analysis on the damage threshold of $MgO:LiNbO_3$ crystals under multiple femtosecond laser pulses," Opt. Mater. 60, 443-449 (2016).
25. M. G. Kuzyk, K. D. Singer, and G. I. Stegeman, "Theory of Molecular Nonlinear Optics," Adv. Opt. Photonics 5, 4 (2013).
26. M. Sheik-Bahae, D. J. Hagan, and E. W. Van Stryland, "Dispersion and band-gap scaling of the electronic Kerr effect in solids associated with two-photon absorption," Phys. Rev. Lett. 65, 96 (1990).
27. W. C. Hurlbut, Y.-S. Lee, K. Vodopyanov, P. Kuo, and M. Fejer, "Multi-photon absorption and nonlinear refraction of GaAs in the mid-infrared," Opt. Lett. 32, 668–670 (2007).
28. M. Sheik-Bahae, A. Said, T.-H. Wei, D. Hagan, and E. Van Stryland, "Sensitive measurement of optical nonlinearities using a single beam," IEEE J. Quantum Electron. 26, 760–769 (1990).
29. L. Mutter, F. D. Brunner, Z. Yang, M. Jazbinsek, and P. Günter, "Linear and nonlinear optical properties of the organic crystal DSTMS," J. Opt. Soc. Am. B 24, 2556–2561 (2007).
30. M. Stillhart, A. Schneider, and P. Günter, "Optical properties of 4-N,N-dimethylamino-4^'-N^'-methyl-stilbazolium 2,4,6-trimethylbenzenesulfonate crystals at terahertz frequencies," J. Opt. Soc. Am. B 25, 1914-1919 (2008).
31. C. Bosshard, R. Spreiter, M. Zgonik, and P. Günter, "Kerr nonlinearity via cascaded optical rectification and the linear electro-optic effect," Phys. Rev. Lett. 74, 2816–2819 (1995).
32. R. DeSalvo, D. J. Hagan, M. Sheik-Bahae, G. Stegeman, E. W. Van Stryland, and H. Vanherzeele, "Self-focusing and self-defocusing by cascaded second-order effects in KTP," Opt. Lett. 17, 28–30 (1992).

33. R. DeSalvo, M. Sheik-Bahae, A. Said, D. J. Hagan, and E. W. Van Stryland, "Z-scan measurements of the anisotropy of nonlinear refraction and absorption in crystals," Opt. Lett. 18, 194–196 (1993).
34. A. Schneider, I. Biaggio, and P. Günter, "Terahertz-induced lensing and its use for the detection of terahertz pulses in a birefringent crystal," Appl. Phys. Lett. 84, 2229–2231 (2004).